# Polarization in the broad lines of NLSy1 and BLSy1 galaxies

**Luka Č. Popović[1]**
*Astronomical Observatoy Belgrade*
*Volgina 7, Belgrade, Serbia*
*E-mail:* `lpopovic@aob.rs`

**Victor L. Afanasiev[2] and Alla I. Shapovalova[2]**
*Special Astrophysical Observatory*
*Special Astrophysical Observatory of the Russian Nizhnij Arkhyz, Karachaevo-Cherkesia 369167, Russia*
*E-mail:* `vafan@sao.ru`

We give an overview of the polarization in the broad lines of Active Galactic Nuclei (AGNs), where we can use the polarization angle for determination of the black hole masses (as given in [1]). Especially, we considered nine AGNs with strong optical Fe II lines in the spectra, where five of them are typical Narrow Line Sy1 (NLSy1) and four Broad Line Sy1 (BLSy1) AGNs.

Comparing the polarization parameters and black hole masses (obtained from polarization) we confirmed that NLSy1 have smaller black hole masses than BLSy1, and seem to have more compact Broad Line Region (BLR) and smaller inner torus radius than BLSy1

---

[1]Luka Č. Popović

1. Introduction – polarization in the AGN broad lines and continuum

It is well known that an Active Galactic Nucleus (AGN) is a powerful source of energy which is produced in the accretion of the gas around a central super-massive black hole. Therefore, to have an AGN one can expect to have a super-massive central black hole, and amount of gas around the black hole. When accretion starts, energy from the central part is emitted, and one can expect emission which can come from X-ray (accretion disk – hot plasma), UV/optical (disk, and broad line region – BLR) to radio emission (mostly from the jets).

Different spectral properties observed in AGN are mostly due to the orientation of the accretion disk and dusty torus toward an observer, that is well explained by, the so called, Unification model (see [2]). Concerning the Unified model one can expect that polarization also strongly depend on the angle of view of an observer. As it is plotted in Fig. 1, observing in the direction of a jet, one can see highly polarized light (in the case of synchrotron radiation of a jet - blazars) or even non-polarized light, while observing in the torus plane direction, it can be seen as polar polarization (scattering on the jet). Observing type 1 AGNs (angle around 45 degrees), one should detect the polarization in the continuum and lines, but the mechanism of polarization is equatorial scattering on the dusty torus (see e.g. [2-5]).

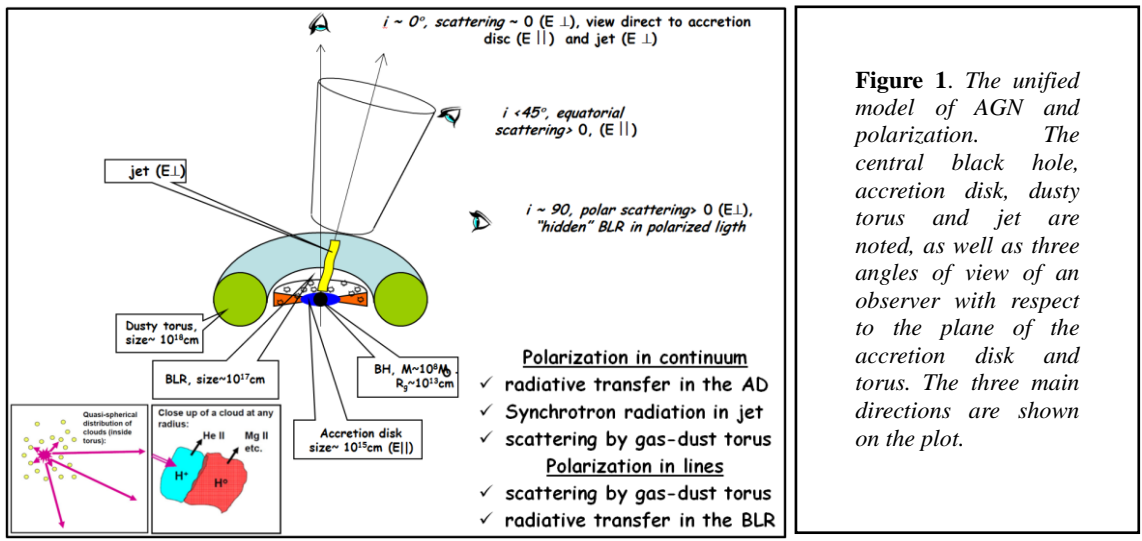

**Figure 1**. *The unified model of AGN and polarization. The central black hole, accretion disk, dusty torus and jet are noted, as well as three angles of view of an observer with respect to the plane of the accretion disk and torus. The three main directions are shown on the plot.*

As it is shown in Fig. 1, the continuum can be polarized in the disk due to the radiative transfer in the accretion disk, and also in the case of synchrotron emission of the jet, while scattering plays an important role in the broad line polarization, since we cannot expect significant contribution of the radiative transfer effects in the BLR. Therefore in the case of broad line polarization (see Fig 1) the orientation is very important. If we observe an AGN in direction parallel to the dusty torus, we can expect that polar scattering will be the dominant effect in polarization, while observing the AGN in the direction around 45 degrees from the dusty torus plane, we expect to have equatorial scattering [3-5]. The orientation is also important in the effect of the continuum polarization. As it can be seen in Fig. 2, we measured the difference between polarization angle of the continuum and the jet projection for a number of Sy1 (AGN with broad lines) and Sy 2 (AGN with only narrow lines) and plot it in Fig. 2. The data for Sy1 are taken from [6-8] and for Sy 2 from [9-11]. It can be clear seen in Fig. 2 that we have a separation between Sy1 and Sy2. As it is expected in the Unification model [2], angles of view to the Sy 1 are between 0 and 45 degrees from the jet and for Sy2 is higher than 45 degrees. It is interesting that we can find both, Sy 1 and Sy 2 (see Fig. 2) with this angle around 45 degrees.

Taking that Sy1 and Sy2 are mostly different because of width of their permited lines, we were motivated to investigate the Narrow Sy1 (NLSy1) galaxies in polarized light to find some significant differences between broad line Sy1 (BLSy1) and NLSy1. Let us recall of some characteristics of NLSy1 in the optical spectra: first of all, their broad lines have Full Width at Half Maximum (FWHM) smaller than 2200 km/s, they have weak [OIII] lines and strong optical Fe II lines.

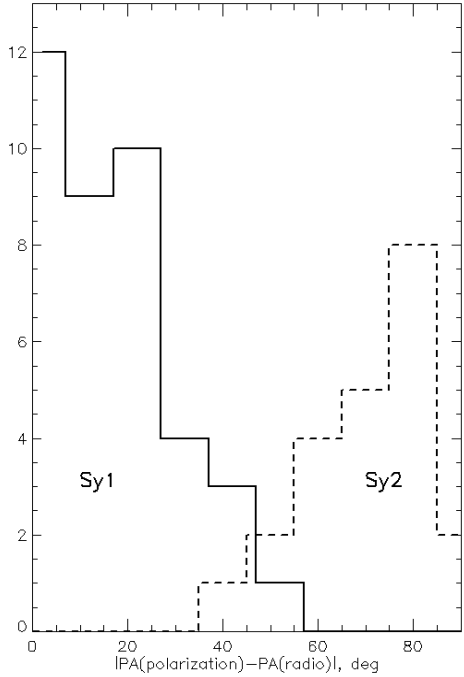

**Figure 2**. *Number of Sy 1 and Sy 2 AGNs as a function of the difference between the polarization angle in the optical continuum and angle of the radio jet. The difference is significantly larger for Sy2 AGNs, that shows the importance of orientation for characteristics of polarized light from AGNs*

Therefore for a sample of 29 AGNs (details can be seen in [12]) we selected five NLSy1(Ark564, NGC 4051 Izw1, Mrk 335, and Mrk 110), and addtional four BLSy1(PG0844+349, IRAS13340+2438, Mrk 1501 and Akn 120) which have larger FWHM (>2200 km/s), but the strong Fe II optical lines and weak [OIII] lines in order to compare some parameters obtained from polarized light. Here we show comparation between these two types of AGNs obtained from observed polarization in the optical spectra.

2. Observations

We performed spectropolarimetric observations of a number of the broad line AGNs with the 6-meter telescope of SAO RAS using the modified spectrograph SCORPIO (see [13] and [14]) in the mode of the spectropolarimetry and polarimetry in the spectral range 4000−8000 Å with the spectral resolution 4 – 10 Å.

As an analyzer we used double analyzer Wollaston - WOLL2 (see [15]). In this analyzer, the beam is divided into two channels, in each channel has been set up a Wollaston prism that divides the direction of the beam at 0-90 and 45-135 degrees. The two obtained images of an object are distributed along the slit using achromatic wedges. In each exposition the four spectra have been registred simultaneously in different polarization planes. In this way we are able to measure three Stokes parameters I, Q and U, from only one observation what is an advantage in comparison with the method that uses only one Wollaston analyzer and consequently one has to change the plate angle and take four expositions.

3.  Results

First we measured central black hole masses in the sample of nine AGNs using the method given in [1]. We found that in principle BLSy 1 have larger masses than NLSy1, the average black hole mass for BLSy1 is log(M/Ms)=8.2, but for NLSy1 is (M/Ms)=7.38 that is almost one order of magnitude smaller than in BLSy1.

Since the measured masses using polarization angle do not depend on the orientation, one can use the virial product $VP=R\sigma/G$, where $R$ is the dimension of the BLR which can be measured from reverberation, $\sigma$ is the dispersion that can be measured from FWHM and G is the gravitational constant. If we have $VP$, then the inclination of the BLR can be obtained using equation

$$\sin^2 i = VP/M_{BH}$$

where $M_{BH}$ is the black hole mass obtained from the polarization angle. We calculated virial product, and use our black hole mass measurements to find the inclination of the BLR in NLSy1 and BLSy1.

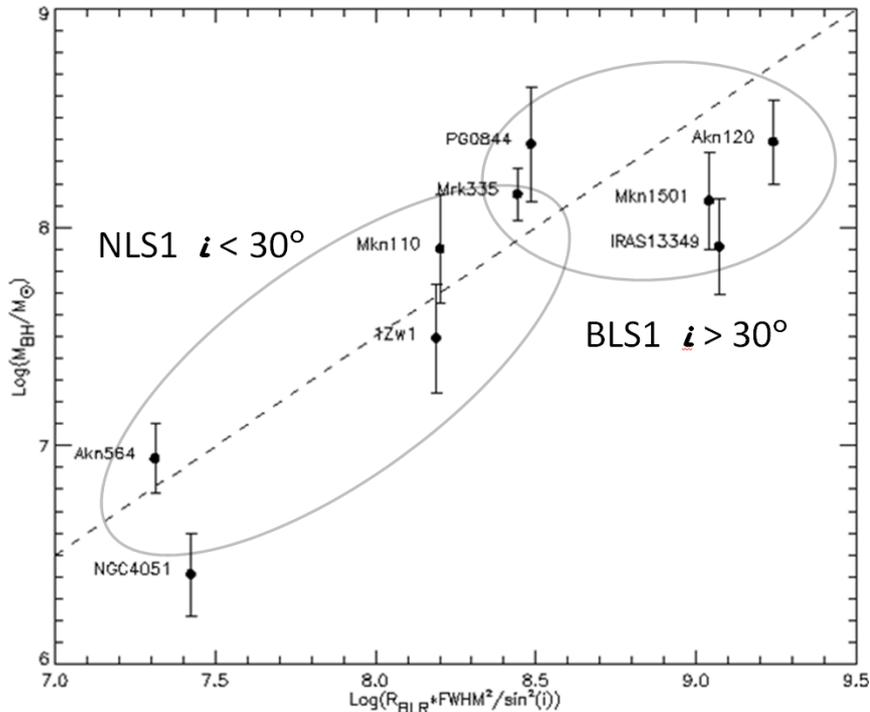

**Figure 3.** *Mass of black hole determined by polarization in the line as a function of VP divided with $\sin^2 i$, the dashed line shows the expected equivalence between masses determined by polarization and one epoch method (assuming virialization and BLR inclination)*

We found that the incliation toward an observer is significantly different. The averaged BLR inclination in NLSy1 is around 24 degrees, but in BLSy1 is almost double, around 40 degrees. It seems that we observe NLSy 1 closer face on than BLSy1. This is shown in Fig. 3

Also, we were able to compare dimensions of the BLR and the radius of the scattering region, and found that dimensions of the BLR in NLSy1 are significantly smaller (in average around 14 light days) than in BLSy1 (in averrage around 46 light days). It is the case also with inner part of the dusty torus, which in NLSy1 is around 98 light days on averrage, but in BLSy1 is on average around 500 light days.

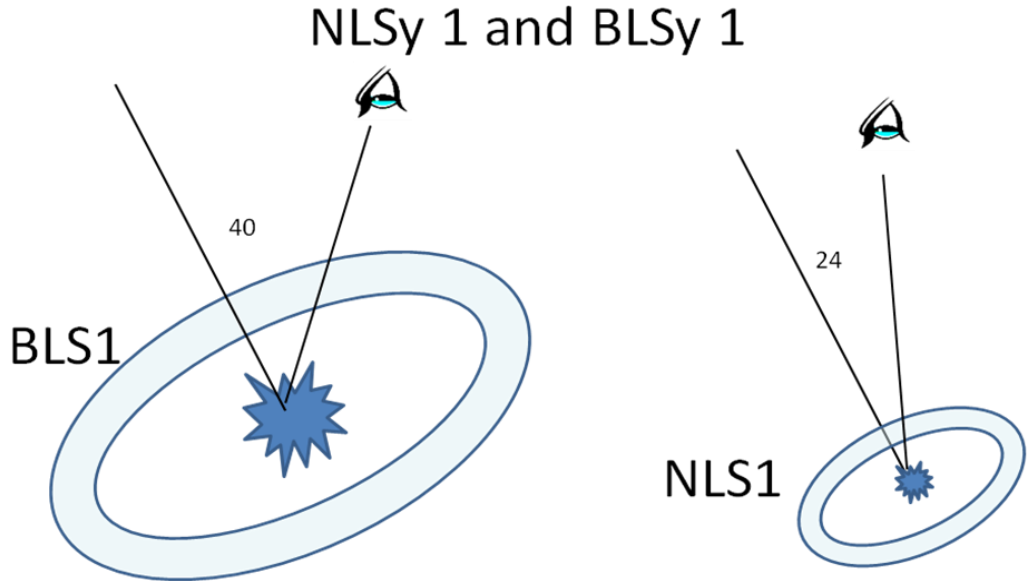

.

**Figure 4**. *Possible difference between NLSy1 and BLSy1 seems to be in dimensions, mass of black hole and in the angle of view*.

Very roughly the picture of differences between NLSy1 and BLSy1 can be presented as it is given in Fig. 4 where a schematic presentation of two types broad line type 1 is given. It seems that these two types are different in dimensions and also in the viewing angle toward an observer.

4. Conclusion

Here we give a short discussion about polarization in NLSy1 AGNs. In order to explore the polariztion in this type of broad line Sy1, we compare the characteristics of a sample of five NLSy1 and four BLSy1. We choose a sample of BLSy1 which have strong Fe II lines and low intensity in [OIII] lines. We found that: a) the black hole masses are more massive in BLSy1 than in NLSy1; b) the dimensions of the BLR and inner part of the dusty torus seem to be significantly smaller in NLSy1 than in BLSy1, and c) The BLR inclination in NLSy1 is around 20 degrees, while in BLSy1 is around 40 degrees, it seems that the angle of view to NLSy1 is almost face-on. This is also in agreement with previous estimated BLR inclinations (see e.g. [16])

All of these results are prelimiarly, and we will investigate in more details these difference in a fortcoming paper [17].

**Acknowledgements**

This work has been financed by Minitry of Education, Science and Technological development of R. Serbia (project 176001) and Russian Foundation for Basic Research (grants N12-02-00857 and N15-02-02101). The work was presented at NLSy1 conference that has been organized with the support of the Department of Physics and Astronomy "Galileo Galilei", the University of Padova, the National Institute of Astrophysics INAF, the Padova Planetarium, and the RadioNet consortium. RadioNet has received funding from the European Union's Horizon 2020 research and innovation programme under grant agreement No 730562.